\documentclass[conference]{IEEEtran}
\usepackage{cite}
\usepackage{graphicx}
\usepackage{array}
\usepackage{tikz}
\usepackage{tikz-3dplot}
\usetikzlibrary{shapes.geometric}
\usetikzlibrary{decorations.pathreplacing}
\usetikzlibrary{positioning}
\usepackage{pgfplots}
\usepackage{pstool}  
\usepackage[top=0.81in, bottom=0.81in, left=0.79in, right=0.79in]{geometry}  
\usepackage[linesnumbered,algoruled,boxed,lined]{algorithm2e}
\usepackage[absolute]{textpos}
\usepackage{threeparttable}
\usepackage[strict]{changepage}
\usepackage[keeplastbox]{flushend} 
\usepackage{bm,upgreek} 
\usepackage{amssymb}
\usepackage{comment}
\usepackage{bbm}
\usepackage{hhline}
\usepackage{amsthm}

\usepackage{lipsum} 
\usepackage{enumitem}

\renewcommand{\thefootnote}{\fnsymbol{footnote}}
\ifCLASSINFOpdf
\else
\fi
\usepackage{amsmath}
\usepackage{amsfonts}
\usepackage{amssymb}

\ifCLASSOPTIONcompsoc
  \usepackage[caption=false,font=normalsize,labelfont=sf,textfont=sf]{subfig}
\else
  \usepackage[caption=false,font=footnotesize]{subfig}
\fi
\usepackage{url}

\newcommand{\Given}{\,\bigg\vert\,}

\pgfplotsset{compat=1.12}
\begin{document}

%

%
\title{\mbox{\hspace*{-0.16in}Deep Q-Learning for Self-Organizing Networks Fault} \mbox{\hspace*{0.05in}Management and Radio Performance Improvement}}
%
%
%

\author{\IEEEauthorblockN{Faris B.~Mismar and Brian L.~Evans}

\IEEEauthorblockA{Wireless Networking and Communications Group, The University of Texas at Austin, Austin, TX 78712 USA}

}

\maketitle

\begin{abstract}
We propose an algorithm to automate fault management in an outdoor cellular network using deep reinforcement learning (RL) against wireless impairments.  This algorithm enables the cellular network cluster to self-heal by allowing RL to learn how to improve the downlink signal to interference plus noise ratio through exploration and exploitation of various alarm corrective actions. The main contributions of this paper are to 1) introduce a deep RL-based fault handling algorithm which self-organizing networks can implement in a polynomial runtime and 2) show that this fault management method can improve the radio link performance in a realistic network setup.  Simulation results show that our proposed algorithm learns an action sequence to clear alarms and improve the performance in the cellular cluster better than existing algorithms, even against the randomness of the network fault occurrences and user movements.
\end{abstract}

\begin{IEEEkeywords}
reinforcement learning, wireless, tuning, optimization, artificial intelligence, SON.
\end{IEEEkeywords}

%
\IEEEpeerreviewmaketitle

\section{Introduction}
%
%
%
%


 

\textit{Self-organizing networks} (SON) are expected to improve the efficiency of cellular coverage tuning to meet target service performance metrics\cite{6963801}.  Industry standards  \cite{3gpp32500} refer to the category of SON which performs automatic handling of network faults as \textit{self-healing}.  Management of network faults is one of the functional areas defined in an industry framework \cite{fcaps} known as FCAPS: fault, configuration, accounting, performance, and security.  Fault management detects faults and corrects them.  Wireless network faults can be corrected through parameter adjustments which optimize the network performance.  This enables it  to carry the traffic it has been dimensioned for with high reliability and end-user \textit{quality of experience} (QoE).  However, this is also a perpetual and costly task.

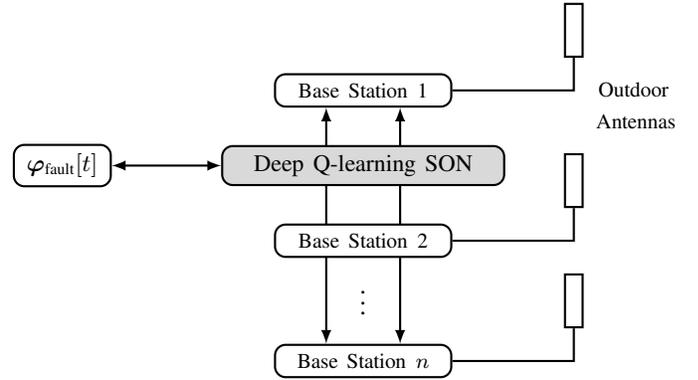
\begin{figure}[!t]
\centering
\begin{tikzpicture}[style=thick,scale=0.8]

\node [rectangle, draw, rounded corners, 
         text width=6em, text centered, minimum height=0.5em] at (4,1) (bs1) {\smaller Base Station 1};
         
\node [rectangle, draw, rounded corners, 
         text width=6em, text centered, minimum height=0.5em] at (4,-3.5) (bs3) {\smaller Base Station $n$};
\node at (4,-2.4) {$\vdots$};

\path [draw, latex-latex] (bs1.335) -- node {} (bs3.25);
\path [draw, latex-latex] (bs1.205) -- node {} (bs3.155);

\node [rectangle, draw, rounded corners, 
         text width=6em, text centered, minimum height=0.5em,fill=white] at (4,-1.5) (bs2) {\smaller Base Station 2};

\node [rectangle, draw, rounded corners, 
            text width=10em, text centered, below of=bs1,fill=gray!30] (fm) {\small Deep Q-learning SON};

\node [rectangle, draw,  minimum width=1, minimum height=2em] at (7.5,2) (ant1) {};
\node [rectangle, draw,  minimum width=1, minimum height=2em] at (7.5,-0.5) (ant2) {};
\node [rectangle, draw,  minimum width=1, minimum height=2em] at (7.5,-2.5) (ant3) {};
     
\node [rectangle, draw, rounded corners, 
    text width=3em, text centered, minimum height=1em, left of=fm, xshift=-3cm](init) {\small $\bm{\varphi}_\text{fault}[t]$};
  
\node[text width=1cm] at (8.5,0.75) {\smaller \centering Outdoor \\ Antennas};

\path [draw, latex-latex] (init.0)  -- node [near start] {} (fm.180);

\draw  (bs1.0) -| node [below right of = ant1, text width=0.5 cm] {} (ant1.270);
\draw  (bs2.0) -| node [below right of = ant2, text width=0.5 cm] {} (ant2.270);
\draw  (bs3.0) -| node [below right of = ant3, text width=0.5 cm] {} (ant3.270);
\end{tikzpicture}%
\vspace*{-0.1in}
\caption{The Deep $Q$-learning self-organizing network (SON) module interacting with several outdoor base stations.}
\label{fig:overall}
\end{figure}

We use deep \textit{reinforcement learning} (RL) where the SON learns fault management with no human supervision.    Our proposed addition of deep RL to the SON is shown in Fig.~\ref{fig:overall}.   SON aided with the deep RL-based algorithm monitors the performance data of an outdoor cellular network and analyzes it to derive proper parameter corrective steps and implements them.  For this purpose, we model a cellular cluster of base stations in an outdoor environment with mobile devices scattered in the vicinity with the focus on the downlink.  We refer to any of these mobile devices as a \textit{user equipment} (UE).  We chose the \textit{signal to interference and noise ratio} (SINR) and throughput as measures of cellular capacity.  We focus on the \textit{fourth generation of wireless communications} or \textit{Long Term Evolution} (4G LTE) and its variants since the system-level simulator \cite{VLS-2016} supports them.  The results can be applied to any similar \textit{orthogonal frequency-division multiplexing} (OFDM) including the \textit{fifth generation of wireless communications new radio} (5G NR) where SON has a highly anticipated role \cite{6963801}.

%
%
%
%

The first deep RL framework to learn control policies using reinforcement learning was introduced in \cite{mnih2013playing}.  This framework outperformed human experts.     The authors in \cite{7393587} used $Q$-learning as part of their SON implementation for mobile load balancing and mobility optimization for cell reselection and handovers.  They used cells with single antennas.  We on the other hand use \textit{multiple-input multiple-output} (MIMO) in our transceivers---a fundamental setup for present and futuristic network deployments.  

Deep learning in mobile and wireless networking with interference alignment was studied in \cite{HeZYZYLZ17}. Relaxed \textit{channel state information} (CSI) assumptions were made in the study where the CSI transition matrix was identical across all users.  We do not make this assumption since we focus on upper layers in the wireless stack.  A two-dimensional convolutional neural network was used in simulations, which imposes unfounded spatially invariant relationships between learning features, or local connection patterns \cite{726791}.  We avoid this in our design of our deep neural network.  The authors in \cite{faris} provided a means to improve the handover execution success rate using supervised machine learning but did not use reinforcement learning which has the ability to learn from previous actions.  In \cite{KHATIB20157549}, a method was proposed to extract the knowledge base from solved fault troubleshooting cases using data mining and supervised learning techniques using fuzzy logic.  Expert opinion was used to define performance measurements and target values.  We on the other hand use RL to derive a policy to map actions to be taken by the self-healing functionality in response to select common number of faults in the network.  
The method in \cite{Dean2012} showed that deep RL can be run in a distributed fashion.

Our main contributions are as follows:
\begin{itemize}
\item Introduce a deep RL-based fault handling algorithm which self-organizing networks can implement in a polynomial runtime.
\item Show that this fault management method can improve the radio link performance in a realistic outdoor network setup.
\end{itemize} 


\section{System Model}\label{sec:sys_mod}
The system comprises a wireless network of a macro base station operating in the sub-6 GHz frequency range with a single tier of surrounding macro base stations, and a machine learning algorithm using deep RL which could  reside in the serving base station or at a central location.

\subsection{Network}\label{sec:network}
The network is an outdoor cellular cluster using frequency division duplex and multiple access with one tier of neighboring cells each with a hexagonal geometry and an inter-site distance of length $L$.  All of these cells are neighbors to one another.  The UEs are equipped with one antenna each and are allowed to perform handovers between cells.   The cells have multiple antennas each.  Any UE is served by a serving cell $c$, which is a member of the set of cells $\mathcal{C}$.  In this network, there are $q\vert\mathcal{C}\vert$ UEs, where $q$ is the number of UEs per cell.

This cellular network can be in a normal operational state or undergo several issues or faults.  The set of faults are $\mathcal{N} = \{\nu_i\}_{i = 0}^{\vert\mathcal{N}\vert-1}$.  We choose a few common faults which can be resolved by SON and set their rate of occurrence as in Table~\ref{table:network_actions}.    

\begin{table*}[!t]
\setlength\doublerulesep{0.5pt}
\caption{Network Faults $\mathcal{N}$}
\label{table:network_actions}
\centering
\begin{threeparttable}
\begin{tabular}{c|lc || c|lc}
\hhline{======}
Action $\nu$ & Definition & Rate & Action $\nu$  & Definition & Rate\\
\hline 
 0 & Cluster is normal. & $p_0$  & 5 & Feeder fault alarm cleared.\tnote{\textdagger} &$p_5$\\
1 & Changed antenna azimuth clockwise (e.g., due to wind). & $p_1$  &  6 & Neighboring cell is up again.\tnote{\textdagger} &$p_6$\\

 2 & Neighboring cell is down. &$p_2$ &  7 & Transmit diversity is normal.\tnote{\textdagger} &$p_7$\\
 3 & Transmit diversity failed. &$p_3$ &  8 & Reset antenna azimuth.\tnote{\textdagger} & $p_8$ \\
 4 & Feeder fault alarm (3 dB loss of signal). & $p_4$ & & \\
 \hhline{======}
\end{tabular}
\begin{tablenotes}\footnotesize
\item[\textdagger] These actions cannot happen if their respective alarm did not happen first.
\end{tablenotes}
\end{threeparttable}
\end{table*}

\subsection{Reinforcement Learning}\label{sec:reinf_learning}

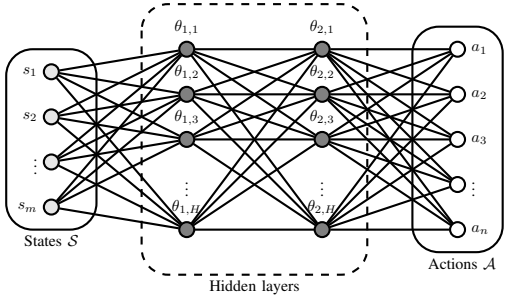
\begin{figure}[!t]
\centering
\begin{tikzpicture}[thick,scale=0.6, every node/.style={scale=0.6},   cnode/.style={draw=black,fill=#1,minimum width=3mm,circle},
]
 \node at (3,-4) {$\vdots$};
  \node at (6,-4) {$\vdots$};
    \foreach \x in {1,...,4}
    {   
    
      \pgfmathparse{\x== 3 ? "\vdots" : "s_\x"}
             \pgfmathparse{\x== 4? "s_m" : "\pgfmathresult"}
        \node[cnode=gray!20,label=180:${\pgfmathresult}$] (x-\x) at (0,{-\x-int(\x/5)-0.5}) {};

        \pgfmathparse{\x<4 ? \x : "H"}
        \node[cnode=gray,label=90:$\theta_{1,\pgfmathresult}$] (x2-\x) at (3,{-\x-int(\x/4)}) {};
        \node[cnode=gray,label=90:$\theta_{2,\pgfmathresult}$] (p-\x) at (6,{-\x-int(\x/4)}) {};
        
    }

	\draw[rounded corners=10pt] (-1,-1) rectangle ++(2,-4) node[below] at (0, -5) {States $\mathcal{S}$};
	\draw[dashed, rounded corners=10pt] (2,0) rectangle ++(5,-6) node[below] at (4.5,-6) {Hidden layers};
	\draw[rounded corners=10pt] (8,-0.5) rectangle ++(2,-5) node[below] at (9.1,-5.5) {Actions $\mathcal{A}$};

	\foreach \x in {1,...,5}
	{
      	\pgfmathparse{\x== 4 ? "\vdots" : "a_\x"}
	\pgfmathparse{\x== 5? "a_n" : "\pgfmathresult"}
        	\node[draw=black,circle,label=0:${\pgfmathresult}$] (s-\x) at (9,{-\x-int(\x/6)+0 }) {};
	}
	    
    \foreach \x in {1,...,4}
    {   \foreach \y in {1,...,5}
        {   \draw (p-\x) -- (s-\y);
        }
    }

    \foreach \x in {1,...,4}
    {   \foreach \y in {1,...,4}
        {  \pgfmathparse{\x<4 ? \x : "H"}   
	        \draw (x2-\x) -- (p-\y) ; 
        }
    }
    \foreach \x in {1,...,4}
    {   \foreach \y in {1,...,4}
        {  \draw (x-\x) -- (x2-\y);
        }
    }
\end{tikzpicture}
\caption{Structure of the neural network used for the DQN implementation with two hidden layers each of dimension $H$.  $\vert\mathcal{S}\vert = m, \vert\mathcal{A}\vert = n$.}
\label{fig:dnn}
\end{figure}

\begin{figure}[!t]
\centering
\begin{adjustwidth}{0.7cm}{0cm}
\begin{tikzpicture}[node distance = 5em, auto, thick, scale=0.9, font=\scriptsize]]
    \node [rectangle, draw, 
    text width=8em, text centered, rounded corners, minimum height=1em,fill=gray!30] (Agent) {Self-Healing Agent};
    \node [rectangle, draw, 
    text width=8em, text centered, rounded corners, minimum height=1em, below of=Agent] (Environment) {Radio Environment};
    
     \path [draw, -latex] (Agent.0) --++ (2em,0em) |- node [text width=5em,near start]{FM Action\\ $a\in\mathcal{A}$} (Environment.0);
     \path [draw, -latex] (Environment.180) --++ (-2em,0em) |- node [text width=5em, xshift=-0.3cm,yshift=-1.3cm] {New state\\ $s^\prime\in\mathcal{S}$} (Agent.180);
     \path [draw, -latex] (Environment.north) --++ (0em,0em) -- node [right] {Reward $r_{s,a}$} (Agent.south);
\end{tikzpicture}
\end{adjustwidth}
\caption{Reinforcement learning elements.  FM stands for fault management.}
\label{fig:reinf}
\end{figure}
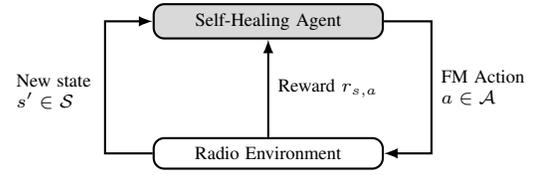

We use $Q$-learning with a deep neural network  as in \cite{mnih2013playing}.  This deep neural network is known as the \textit{deep $Q$-network} (DQN) and is shown in Fig.~\ref{fig:dnn}.  We use DQN with experience replay and formulate the SON fault management as a deep RL problem as in Algorithm~\ref{alg:the_fm_alg}.  The framing of the reinforcement learning problem is shown in Fig.~\ref{fig:reinf}.  

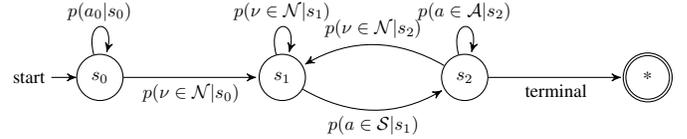
\begin{figure}[!t]
\centering
\usetikzlibrary{automata,positioning}
    \begin{tikzpicture}[->,>=stealth',scale=0.7, every node/.style={scale=0.7,node distance=1.8cm}]
        \node[initial,state] (s0) {$s_0$};
        \node[state, right=of s0] (s1) {$s_1$};	
        \node[state, right=of s1] (s2) {$s_2$};	
        \node[accepting,state, right=of s2] (e0) {*}; 
        \draw[every loop]
(s0) edge [loop above] node {$p(a_0\vert s_0)$} (s0)
(s1) edge[bend right, auto=right] node {$p(a \in \mathcal{S}\vert s_1)$} (s2)
(s1) edge[loop above] node {$p(\nu\in\mathcal{N} \vert s_1)$} (s1)
(s2) edge[loop above] node {$p(a \in\mathcal{A}\vert s_2)$} (s2)
(s2) edge[bend right, auto=right] node {$p(\nu\in\mathcal{N}\vert s_2)$} (s1)
(s2) edge[right, auto=right] node {terminal} (e0)
(s0) edge[right, auto=right] node {$p(\nu\in\mathcal{N} \vert s_0)$} (s1)
	    ;
    \end{tikzpicture}
    \caption{Markov decision process used in the formulation of the SON fault handling algorithm.}
    \label{fig:markov_volte}
\end{figure}

To formulate this problem as a reinforcement learning problem, we define a \textit{Markov decision process} (MDP) with transition probabilities $p(\cdot)$ as in Fig.~\ref{fig:markov_volte}.  The set of actions carried out by the self-healing agent are $\mathcal{A} = \{a_i\}_{i = 0}^{n-1}$ and the set of network states are $\mathcal{S} = \{s_i\}_{i = 0}^{m-1}$.  These are shown in Table~\ref{table:simulated_son_actions}.    A state $s$ is \textit{terminal} if it is the final state or if the objective has been met.  We define the reward as:
\begin{equation}
r_{s,a}[t] \triangleq \begin{cases} 
	r_1, & \; 0 < \vert{\bm\varphi}_\text{fault}[t - 1]\vert < \vert{\bm\varphi}_\text{fault}[t]\vert  \\
	      r_2, & \;  \vert{\bm\varphi}_\text{fault}[t]\vert = \vert{\bm\varphi}_\text{fault}[t-1] \vert  \\
      r_3, &\;  \vert{\bm\varphi}_\text{fault}[t] \vert< \vert{\bm\varphi}_\text{fault}[t-1]\vert  \\
r_4, & \;  \vert{\bm\varphi}_\text{fault}[t]\vert = 0  \, \text{(i.e., objective is met)}
   \end{cases}
   \label{eq:rewards}
\end{equation}
where $\vert\varphi_\text{fault}[t]\vert$ is the number of bits that are set to logic-1 in the fault register at a given TTI $t$. The fault register $\varphi_\text{fault}[t]$ is a register of $u$ boolean entries, where the $i$-th entry in the register corresponds to the fault with identifier $i$ triggered in this cluster.  It is initialized to all logic-0 and set whenever a fault $i$ happens in the network and unset only when all similar faults are cleared.  These faults are shown in Table~\ref{table:network_actions}. 

With a network having $\vert\mathcal{C}\vert$ cells, $n$ alarm-clearing actions, and $m$ states, the number of elements in a table required is in $\mathcal{O}(mn \vert\mathcal{C}\vert)$.  In networks with thousands of cells and alarms, the tabular $Q$-learning method to keep track of the state-action values in a table does not scale due to this multiplicative nature.

The \textit{episode} $z\colon z \in \{0,1,\ldots,\zeta\}$ is a period of time for an agent-environment interaction to take place. During an episode, the deep RL agent gathers \textit{experience} from a finite number of episodes where each episode has a duration of $\tau$ \textit{transmit time intervals} (TTIs).  This agent stores the experience $e[t]$  defined as $(s, a, r_{s,a}, s^\prime)$   in  a dataset called the \textit{replay memory} $\mathcal{D}$ \cite{mnih2013playing}. 

We next define the estimated function $Q^*(s,a)$, which is the expected discounted reward when starting in state $s$ and selecting an action $a$  as \cite{mnih2013playing}:

\begin{equation}
    \label{eq:bellman_deep}
    Q^*(s,a) \triangleq \mathbb{E}_{s^\prime} \left [ r_{s,a} + \gamma \max_{a^\prime} Q^*(s^\prime,a^\prime) \Given s, a \right ]
\end{equation}

If we define a DQN with its weights at time $t$ as ${\bm\theta}_t$, then (\ref{eq:bellman_deep}) can be approximated as $Q(s,a;{\bm\theta}_t) \approx Q^*(s,a)$ as $t\to \infty$.  This DQN is trained through minimizing the \textit{mean squared error} (MSE) convex loss function
\begin{equation}
    \label{eq:loss}
    L_t({\bm\theta}_t) \triangleq \mathbb{E}_{s,a} \left [(y_t - Q(s,a;{\bm\theta}_t))^2 \right ]
\end{equation}
where $y_t$ is an estimate obtained from the DQN using its weights at time $t-1$ as:
\begin{equation}
    \label{eq:yest}
    y_t \triangleq \mathbb{E}_{s^\prime} \left [ r_{s,a} + \gamma \max_{a^\prime} Q(s^\prime,a^\prime;{\bm\theta}_{t-1}) \Given s, a \right ].
\end{equation}
The weights ${\bm\theta}$ are updated after every iteration in time using a method of the stochastic gradient descent algorithm (SGD) called ``adaptive moments'' \cite{KingmaB14}.  We also use the \textit{rectified linear unit} (ReLU) $x\mapsto \max(0,x)$ as the activation function of each node in the DQN.  This deep learning process repeats for all the episodes $z$.


\begin{table*}[!t]
\setlength\doublerulesep{0.5pt}
\caption{SON Fault Management Algorithm -- Simulated Actions $\mathcal{A}$ and States $\mathcal{S}$}
\label{table:simulated_son_actions}
\vspace*{-.1in}
\centering
\begin{tabular}{ clcl } 
\hhline{====}
Action $a$ & Definition & State $s$ & Definition \\
\hline 
0 & No actions issued. & 0 & No actions issued (a transient state). \\
1 & Faulty neighbor cell is up again. & 1 & Number of active alarms has increased. \\
2 & Serving cell transmit diversity enabled.  & 2 & Number of active alarms has decreased. \\
3 & Serving cell losses recovered.  \\
4 & Serving cell azimuth set to default value.\\
\hhline{====}
\end{tabular}
\end{table*}


During an episode $z$, the deep RL agent tries to maximize the total value of the discounted rewards it receives in response to its action.  It uses a near-greedy action selection rule.  This is because with large number of episodes $\zeta$, every actions will have been sampled enough for the state-action value function to converge \cite{Sutton}. The near-greedy action selection rule of choice is the $\epsilon$-greedy strategy for learning. In this strategy, $\epsilon$ is the \textit{exploration rate} and serves to select a random action $a\in\mathcal{A}$ with a probability $\epsilon\colon 0 < \epsilon < 1 $, known as exploration, or selects an action from the replay memory $\mathcal{D}$, which is also known as exploitation, with a probability $1-\epsilon$. The exploration rate decays in every episode until it reaches the minimum exploration rate $\epsilon_\text{min}$.

 
\renewcommand{\thefootnote}{\fnsymbol{footnote}}


\section{Fault Handling Algorithms}\label{sec:algorithm}
%
%

\subsection{Random}
The faults are cleared from a random sample of active faults.  We choose the discrete uniform random distribution for the healing of the faults in the network since the discrete uniform distribution maximizes the discrete entropy \cite{Cover}.  This is a non-trivial lower bound of performance compared to taking no action.

\subsection{First-In, First-Out}
In this approach, the SON takes fault handling actions  immediately in the next TTI in the order these faults happen.


\subsection{Proposed}
Our proposed deep RL-based algorithm  is shown in Algorithm~\ref{alg:the_fm_alg}. Unlike FIFO, this algorithm can handle faults regardless of the order they happen due to the $\epsilon$-greedy learning strategy.

\begin{algorithm}[!t]
 \small
  \caption{\small SON Fault Management}
  \label{alg:the_fm_alg}
 \DontPrintSemicolon
  \KwIn{The set of fault handling actions $\mathcal{A}$ in a network $\mathcal{C}$.}
  \KwOut{Optimal fault handling commands given during a frame $z$, which has a duration of $\tau$.}
  Define the fault management states $\mathcal{S}$, the exploration rate $\epsilon$, the decay rate $d$, the discount factor $\gamma$, and minimum exploration rate $\epsilon_\text{min}$.\; 
  Initialize time, states, actions, fault handling register, and replay memory $\mathcal{D}$.\;
  \Repeat {$\vert\bm{\varphi}_\text{fault}[t]\vert = 0$ {\rm or} $t \ge\tau$} {
  $t := t  + 1$\;
  $\epsilon := \max(\epsilon\cdot d, \epsilon_\text{min})$ \;
    Sample $r \sim \text{Uniform}(0,1)$\;
  \eIf {$r \le \epsilon$} {
  Select an action $a \in \mathcal{A}$ at random.\;
  } {
  Select an action $a = \arg\max_{a^\prime} Q(s,a^\prime;\bm{\theta}_t)$. \;
  }
  Perform action $a$ to resolve alarm and update $\bm{\varphi}_\text{fault}[t]$.\;
  Obtain reward $r_{s,a}$ from (\ref{eq:rewards}).\;
  Observe next state $s^\prime$.\;
Store experience $e[t] \triangleq (s, a, r_{s,a}, s^\prime)$ in $\mathcal{D}$.\;
Sample from $\mathcal{D}$ for experience $e_j\triangleq (s_j, a_j, r_j, s_{j+1})$.\; 
\eIf {$s_{j+1}$ {\rm is terminal}} {
Set $y_j:= r_j$ 
} {
Set $y_j := r_j + \gamma\max_{a^\prime} Q(s_{j+1}, a^\prime; \bm{\theta}_t)$ 
}
 Perform SGD on $(y_j - Q(s_j, a_j; \bm{\theta}_t))^2$\; 
  $s := s^\prime$\;
}
  Proceed to the next LTE-A frame $z+1$.\;
 \end{algorithm}
 
We next compute the time complexities of these various fault handling algorithms:
\begin{itemize}
\item For the random algorithm, an action is randomly sampled from a list of actions, therefore it has a time complexity in $\mathcal{O}(1)$ per iteration or $\mathcal{O}(\tau)$ total. 
\item The First-In First-Out (FIFO) fault-handling algorithm reviews the alarm register every TTI and therefore has a time complexity in $\mathcal{O}(\max(u, \vert\mathcal{C}\vert))$ per iteration or $\mathcal{O}(\tau \max(u, \vert\mathcal{C}\vert))$ total.  
\item For our proposed algorithm, the time complexity of the for DQN backpropagation algorithm is at least in $\mathcal{O}(k(\bm{\theta})\tau\zeta \vert\mathcal{C}\vert\vert\mathcal{A}\vert)$ \cite{scikit-learn}, where $k(\bm{\theta})$ is a function of the depth and number of the hidden layers $\bm{\theta}$.   
\end{itemize}

Although our proposed algorithm has the highest time complexity cost, the complexity is not dependent on the number of UEs being served, and therefore it is scalable in the number of UEs served in a cluster. Furthermore, while the random algorithm can be trivially distributed across multiple cells independently and require a space complexity in $\mathcal{O}(\vert\mathcal{A}\vert)$, the distributed implementation is less scalable for the FIFO model since it requires $\vert\mathcal{C}\vert(\vert\mathcal{C}\vert - 1)/ 2$ links to communicate about faults and fault management leaving us with a message passing complexity in $\mathcal{O}(\vert\mathcal{C}\vert^2)$.  Our proposed algorithm can be run in a distributed fashion owed to its deep learning component \cite{Dean2012} and centralized collection point at the SON leaving us with message passing complexity in $\mathcal{O}(\vert\mathcal{C}\vert)$.  We refer to the source code \cite{mycode} for further implementation details. 
 
\section{Performance Metrics}\label{sec:benchmarks}
We evaluate the downlink SINR with power allocation using the waterfilling algorithm at the transmitter and the zero-forcing equalization at the receiver as in \cite{VLS-2016}.
We also use the average downlink cell throughput and the downlink user throughput, which is derived from its empirical cumulative distribution function (CDF) as follows: peak (95\%), average, and edge (5\%) \cite{link}.
\begin{table}[!t]
\centering
\caption{Machine Learning Parameters}
\label{table:ml_parameters}
\setlength\doublerulesep{0.5pt}
\vspace*{-0.1in}
\begin{tabular}{ lr } 
\hhline{==}
Parameter & Value \\
\hline
Neural network hidden layer width $H$ & $24$ \\
Loss function $L_t({\bm\theta}_t)$ & mean squared error \\
Optimizer &  \cite{KingmaB14} \\
Number of episodes $\zeta$ & 50  \\
Discount factor $\gamma$ & 0.950 \\
Exploration rate $\epsilon$ & 1.000\\
Exploration rate decay $d$ &  0.91 \\
Minimum exploration rate $\epsilon_\text{min}$ & 0.010\\
One episode duration $\tau$ & 20 TTIs \\
Number of states & 3 \\
Number of actions & 5 \\
\hhline{==}
\end{tabular}
\end{table}

\section{Simulation Results}\label{sec:results}

\begin{table*}[!t]
\setlength\doublerulesep{0.5pt}
\caption{Radio Environment Parameters}
\label{table:parameters}
\centering
\begin{threeparttable}
\vspace*{-0.2in}
\begin{tabular}{ lrlr}
\hhline{====}
Parameter & Value & Parameter & Value\\
 \hline
Bandwidth  & 10 MHz & Downlink center frequency & 2100 MHz \\
LTE cyclic prefix & Normal & Cellular geometry & Hexagonal \\
Inter-site distance &  200m & Scheduling algorithm & Proportional Fair \\
Number of cells in the network & 21 & Propagation model & COST231 \\
Propagation environment & Urban & Number of concurrently active UEs per cell $q$\tnote{\textdagger}  & $ \{10  ,50\}$ \\
BS  antenna model\tnote{\textdagger}  & 3gpp 36.942 & BS maximum transmit power & 46 dBm \\
BS antenna height & 25 m & BS antenna electrical tilt & 4$^\circ$ \\
BS MIMO configuration (\# Tx, \# Rx antennas) & $(4,2)$ & Noise power density & -174 dBm/Hz \\
UE average movement speed & 3 km/h   & UE height & 1.5 m \\
Shadow fading margin standard deviation & 8 dB& BS number of sectors per site & 3 \\
\hhline{====}
\end{tabular}
\begin{tablenotes}\footnotesize
\item[\textdagger]  BS is short for \textit{base station} and UE is short for \textit{user equipment}.
\end{tablenotes}
\end{threeparttable}
\end{table*}


\begin{table*}[!t]
\setlength\doublerulesep{0.5pt}
\caption{Throughput performance per algorithm for different number of users per cell}
\label{table:dl_ue_throughput}
\vspace*{-0.1in}
\centering
\begin{tabular}{ p{0.1\textwidth}ccc|ccc|ccc || ccc} 
 & \multicolumn{9}{c}{User Equipment (UE) Throughput [Mbps]}  & \multicolumn{3}{c}{Cell Throughput [Mbps]} \\
 \cline{2-13}
  & \multicolumn{3}{c}{Random\tnote{\textdagger}} & \multicolumn{3}{c}{FIFO} & \multicolumn{3}{c}{Proposed} & Random & FIFO & Proposed  \\
\cline{2-13}
UEs per cell & Peak &  Average & Edge  & Peak & Average & Edge  & Peak & Average & Edge & \multicolumn{3}{c}{Average}\\
\hline
$q = 10$ & 3.48 & 1.78 & 0.53 & {3.52} & {1.79} & {0.54} & \textbf{3.55} & \textbf{1.84} & \textbf{0.58} & 17.77 &  17.95 & \textbf{18.37} \\
$q = 50$ & 0.68 & 0.38 & 0.13 & 0.68 & 0.38 & 0.13  & 0.68 & 0.38 & 0.13 & 18.81 & 18.96 & \textbf{18.97}  \\
\hhline{=============}
\end{tabular}
\end{table*}

\begin{figure}[!t]
\centering
\scalebox{0.36}{\input{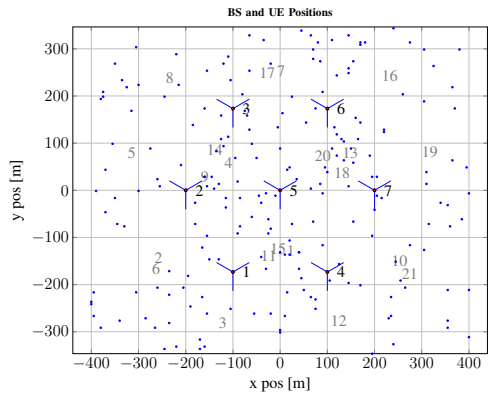}}
\vspace*{-.1in}
\caption{The simulated outdoor network.}
\label{fig:network}
\end{figure}

We use the Vienna LTE-A Downlink System Level Simulator to simulate the outdoor network in Fig.~\ref{fig:network} with reproducibility  \cite{VLS-2016}. The choice between \mbox{LTE-A} or 5G in the sub-6 GHz range is driven by the OFDM numerology scaling factor.  We implement Algorithm~\ref{alg:the_fm_alg} using both MATLAB and Python \cite{mycode}.  The parameters used in our simulation are summarized in~\mbox{Tables}~\ref{table:ml_parameters}~and~\ref{table:parameters}.   Each episode has a duration of $\tau=50$ TTIs.  In LTE-A or 5G, the duration of 1 TTI is equal to 1 ms.   We set the rate of occurrences of faults in Table~\ref{table:network_actions} as $p_0 = 5/9, p_1 = p_2 = p_3 =p_4 = 1/9$.  This way we give all faults an equally likely chance of occurrence which is the worst case.  For the rewards in (\ref{eq:rewards}), we set $r_1 = -1, r_2 = 0, r_3 = 1, r_4 = 5$.  

\begin{figure}[!t]
\centering
\scalebox{0.36}{\input{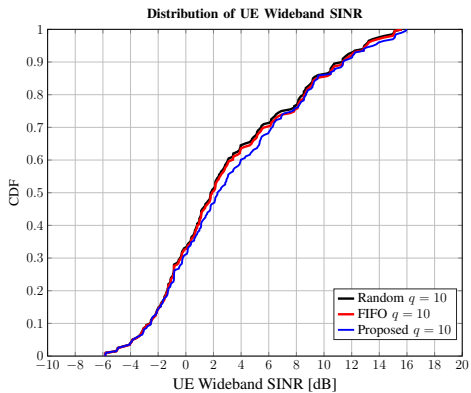}}
\vspace*{-0.1in}
\caption{The average downlink SINR empirical CDF as measured by the user equipment (UEs) for all three algorithms ($q = 10$).}\label{fig:sinrcdf}
\end{figure}

In Fig.~\ref{fig:sinrcdf}, we show the average downlink SINR distribution for all three algorithms for $q=10$.  While the FIFO algorithm performs incrementally better than the random algorithm, our proposed algorithm has a slightly higher downlink SINR overall and a significantly higher SINR gain in the range from 2 to 7 dB.  The random algorithm leads to the worst performance due to the poor handling of faults. For our proposed algorithm, we show the results after $\zeta= 50$ episodes, which is after a total of $\tau\zeta =1$ second. 

The quantitative results of our simulation are shown in Table~\ref{table:dl_ue_throughput}. We observe that when the cell serves a low number of users (i.e., low cell load), our proposed algorithm outperforms the random algorithm as a lower bound and outperforms the FIFO algorithm. However, as the cell serves more UEs ($q=50$), the performance of all algorithms becomes similar since the cellular resources are near depleted (i.e., a bottleneck) at highload and therefore clearing alarms does not lead to significant improvements.

\section{Conclusion}\label{sec:conclusions}

In this paper, we attempted to find an optimal solution policy for a RL based fault management algorithm in an outdoor realistic cellular environment.  We motivated the need for deep RL in resolving  faults in similar cellular environments.
The use of deep RL allows the distributed implementation of the fault handling algorithm, which is useful when the cluster size is large.
Our proposed solution works by allowing RL to learn how to improve the DL SINR through exploration and exploitation of various alarm corrective actions.
The simulations showed that the deep RL-based method, which we proposed, can improve the performance of the cellular network measured by the downlink SINR and downlink throughputs.
Therefore, the proposed deep RL-based automated cellular network tuning framework is useful in maintaining the end-user QoE in a network with impairments and faults.
 



%


\ifCLASSOPTIONcaptionsoff
  \newpage
\fi



%

\bibliography{references}  

\bibliographystyle{IEEEtran}

%









\end{document}